\documentclass{aastex631}

\newcommand{\beq}{\begin{equation}}
\newcommand{\eeq}{\end{equation}}

\newcommand{\Ms}{\textrm{M}_*}
\newcommand{\Msun}{\textrm{M}_\odot}
\newcommand{\kmps}{km~s$^{-1}$}
\newcommand{\hi}{H{\sc i}}

\newcommand{\hii}{H{\sc i}\,21cm}

\newcommand{\Mmol}{\rm{M_{mol}}}
\newcommand{\cii}{[C{\textsc{ii}}]~158$\mu$m}

\begin{document}

\title{An H{\sc i}-absorption-selected cold rotating disk galaxy at $z\approx2.193$ }

\author{B. Kaur} 
\affiliation{National Centre for Radio Astrophysics, Tata Institute of Fundamental Research, 
Pune University, Pune 411007, India}

\author{N. Kanekar} 
\affiliation{National Centre for Radio Astrophysics, Tata Institute of Fundamental Research, Pune University, Pune 411007, India}

\author{M. Neeleman}
\affiliation{National Radio Astronomy Observatory, 520 Edgemont Road, Charlottesville, VA 22903, USA}

\author{M. Rafelski}
\affiliation{Space Telescope Science Institute, 3700 San Martin Drive, Baltimore, MD 21218, USA}
\affiliation{Department of Physics \& Astronomy, Johns Hopkins University, Baltimore, MD 21218, USA}

\author{J. X. Prochaska}
\affiliation{Department of Astronomy \& Astrophysics, UCO/Lick Observatory, University of California, 1156 High Street, Santa Cruz, CA 95064, USA}
\affiliation{Kavli Institute for the Physics and Mathematics of the Universe (Kavli IPMU), 5-1-5 Kashiwanoha, Kashiwa, 277-8583, Japan}

\author{R. Dutta}
\affiliation{Inter-University Centre for Astronomy and Astrophysics, Pune University, Pune 411007, India}

\begin{abstract}
We have used the Atacama Large Millimeter/submillimeter Array (ALMA) to map CO(3--2) emission from a galaxy, DLA-B1228g, associated with the high-metallicity damped Lyman-$\alpha$ absorber at $z \approx 2.1929$ towards the QSO PKS~B1228-113. At an angular resolution of $\approx0\farcs32\times0\farcs24$, DLA-B1228g shows extended CO(3--2) emission with a deconvolved size of $\approx0\farcs78\times0\farcs18$, i.e. a spatial extent of $\approx6.4$~kpc. We detect extended stellar emission from DLA-B1228g in a Hubble Space Telescope Wide Field Camera~3 F160W image, and find that H$\alpha$ emission is detected in a Very Large Telescope SINFONI image from only one side of the galaxy. While the clumpy nature of the F160W emission and the offset between the kinematic and physical centers of the CO(3--2) emission are consistent with a merger scenario, this appears unlikely due to the lack of strong H$\alpha$ emission, the symmetric double-peaked CO(3--2) line profile, the high molecular gas depletion timescale, and the similar velocity dispersions in the two halves of the CO(3--2) image. Kinematic modelling reveals that the CO(3--2) emission is consistent with arising from an axisymmetric rotating disk, with an exponential profile, a rotation velocity of $v_{rot}=328\pm7$~km~s$^{-1}$, and a velocity dispersion of $\sigma_{v}=62\pm7$~km~s$^{-1}$. The high value of the ratio $v_{rot}/\sigma_v$, $\approx5.3$, implies that DLA-B1228g is a rotation-dominated cold disk galaxy, the second case of a high-$z$ H{\sc{i}}-absorption-selected galaxy identified with a cold rotating disk. We obtain a dynamical mass of $M_{dyn}= (1.5\pm0.1)\times10^{11}~M_\odot$, similar to the molecular gas mass of $\approx10^{11}~M_\odot$ inferred from earlier CO(1--0) studies; this implies that the galaxy is baryon-dominated in its inner regions.

\end{abstract}


\section{Introduction} \label{sec:intro}

Damped Ly$\alpha$ absorbers (DLAs), the highest \hi\ column density absorbers in QSO absorption spectra, provide a unique way of identifying gas-rich galaxies at high redshifts \citep[e.g.][]{Wolfe05}. In the local Universe, DLA column densities, N$_{\rm HI} \geq 2 \times 10^{20}$~cm$^{-2}$, arise in or around galaxies (e.g. in galaxy disks, high-velocity clouds, intra-group gas, etc); the presence of a DLA along a QSO sightline has hence long been used to identify high-$z$ galaxies via the \hi\ absorption signature \citep[e.g.][]{Wolfe86}. The importance of such ``\hi-absorption-selected galaxies'' stems from the fact that their selection is not biased towards high-luminosity objects. In contrast, emission-selected galaxy samples implicitly pick out the brighter galaxies of the population. 
While studies of emission-selected samples have provided a wealth of information on the properties of high-redshift galaxies \citep[e.g.,][]{Madau14,Tacconi20}, the luminosity bias inherent in the selection can affect the interpretation of the results. Addressing this issue critically requires the detection and detailed characterization of the \hi-absorption-selected galaxies associated with high-redshift DLAs.

For more than three decades, a range of approaches, including broad-band and narrow-band imaging, multi-angle long-slit spectroscopy, and integral field spectroscopy, have been employed at optical and near-IR wavelengths to search for the galaxies associated with high-$z$ DLAs \citep[e.g.,][]{Moller93,Warren01,Kulkarni06,Peroux12,Fynbo10,Bouche12,Fumagalli14,Wang15,Krogager17,Lofthouse23,Oyarzun24}. 
Unfortunately, the success of these studies has been limited by the faintness of the stellar emission of the DLA galaxies.
Despite the large number of searches, only $\approx 25$ \hi-absorption-selected galaxies have been identified via these methods at $z \gtrsim 2$ \citep[e.g.][]{Krogager17}, some at very large impact parameters ($> 100$~kpc) to the QSO sightline \citep[e.g.][]{Lofthouse23}.

In the last few years, mm- and sub-mm searches for the galaxy counterparts of high-$z$ DLAs with the Atacama Large Millimeter/submillimeter Array (ALMA) and the Northern Extended Millimetre Array (NOEMA) have changed the field. ALMA and NOEMA detections of CO and \cii\ emission in the fields of high-metallicity DLAs have yielded the identification of more than 15 \hi-absorption-selected galaxies out to $z \approx 4.3$ \citep[e.g.][]{Neeleman17,Neeleman18,Neeleman19,Kanekar20,Kaur22c,Combes23}. 
At $z \approx 2$, the galaxies identified via CO emission have been found to have relatively low impact parameters ($b \approx 5-30$~kpc, with a median value of $\approx 11$~kpc) to the QSO sightline, or to have a companion at a lower impact parameter; this suggests that the Ly$\alpha$ absorption typically arises in the interstellar medium (ISM) of the DLA galaxy \citep{Neeleman18,Fynbo18,Kanekar20,Kaur22c}. Conversely, at $z \approx 4$, objects identified using \cii\ emission have higher impact parameters, $b \approx 15-60$~kpc (with a median value of $\approx 27$~kpc), suggesting the presence of high-\hi\ column density gas in the circumgalactic medium (CGM) of galaxies at these redshifts \citep[][]{Neeleman17, Neeleman19}. Surprisingly, a number of the newly-identified \hi-absorption-selected galaxies, associated with high-metallicity DLAs, have been found to have high molecular gas masses, $\gtrsim 5 \times 10^{10} \, \Msun$ \citep[e.g.][]{Neeleman18,Neeleman20,Kanekar20,Combes23}. Follow-up ALMA and Karl G. Jansky Very Large Array (VLA) studies have revealed that these high-$z$ DLA galaxies also have high, near-thermal, excitation of the mid-J CO rotational levels \citep{Klitsch22,Kaur22b}. This suggests that a significant fraction of high-metallicity DLAs at $z \approx 2$ arise in the ISM or CGM of massive, dusty, star-forming galaxies.

Recently, H$\alpha$, CO, and \cii\ mapping studies, mostly of emission-selected galaxy samples, have revealed the presence of rotating disk galaxies out to $z \approx 4.5$ \citep[e.g.][]{Bouche13,Genzel17,Neeleman20,Rizzo20,Lelli21}. Interestingly, some of these high-$z$ rotating disks are ``cold'', i.e. their kinematics are dominated by rotation. In the case of \hi-absorption-selected galaxies, ALMA mapping studies have so far been carried out for two systems at $z \approx 4$, using the \cii\ line, finding a major merger for the $z \approx 3.7978$ DLA  towards J1201+2117 \citep{Prochaska19} and a cold rotating disk galaxy for the $z \approx 4.2546$ DLA towards J0817+1351 \citep[the ``Wolfe disk''; ][]{Neeleman20}. In this {\it Letter}, we present the first case of high-resolution CO mapping of a high-redshift \hi-absorption-selected galaxy, at $z \approx 2.193$ towards the quasar PKS~B1228-113.\footnote{We use a flat $\Lambda$ cold dark matter cosmology, with $\Omega_{\Lambda} = 0.7$, $\Omega_{m} = 0.3$, and H$_0 = 70$~\kmps~Mpc$^{-1}$, throughout the paper.}

\section{ALMA Observations and Data Analysis} 
\label{sec:obs}

We used the ALMA Band-3 receivers to observe the redshifted CO(3--2) line from the $z \approx 2.193$ DLA galaxy towards PKS~B1228-113. This DLA field was originally searched for CO(3--2) emission by \citet{Neeleman18}, as part of a sample of high-metallicity DLAs 
\citep{Kanekar20}, and yielded a detection of strong CO(3--2) emission at low angular resolution ($2\farcs6 \times 1\farcs7$).
The new ALMA observations were carried out in July 2021 (ALMA proposal ID~2019.1.00373.S; PI M. Neeleman), using the extended C43-7 configuration (longest baseline: 3.6~km) with an on-source integration time of 4.1~hours. PKS~B1228-113 was chosen as the pointing centre as it is bright at mm wavelengths, allowing self-calibration of the target field. One of the four ALMA basebands was centred on the redshifted CO(3--2) line frequency of the $z = 2.1929$ DLA \citep[$\nu_{obs} = 108.291$~GHz;][]{Neeleman18}, with a bandwidth of 1.875~GHz, 960~channels, and the frequency division mode, yielding a frequency resolution of $\approx 1.95$~MHz. The remaining three basebands were set up in the time division mode, each with a bandwidth of 2.0~GHz and 128 channels, giving a frequency resolution of 15.625~MHz.

We used the standard ALMA pipeline in the Common Astronomy Software Application package \citep[{\sc casa} version~6.5.4.9;][]{casa22,Hunter23} to carry out the initial data editing and calibration of the visibility dataset. After inspecting the calibrated visibilities and performing manual flagging, we  averaged the data to a coarse spectral resolution for the purpose of self-calibration. The self-calibration process involved three rounds of continuum imaging and phase-only self-calibration, followed by one round of amplitude-and-phase self-calibration and imaging, after which no further improvement was seen in either the image or the image/visibility residuals upon further self-calibration. In the self-calibration procedure, the antenna-based gains were measured using the routine {\sc gaincalR} \citep{Chowdhury20}. The continuum imaging in the self-calibration procedure used Briggs weighting \citep{Briggs95} with a robust value of $-0.5$. For the final continuum image, after self-calibration, we used a robust value of $+0.5$, obtaining a synthesized beam of $0\farcs29 \times 0\farcs22$ and an RMS noise of 15~$\mu$Jy~beam$^{-1}$. PKS~B1228-113 has a flux density of $24.85 \pm 0.02$~mJy in this image.

We next applied the antenna-based gains obtained from the self-calibration procedure to the CO(3--2) visibility dataset at the original spectral resolution, and subtracted out the continuum emission from the calibrated visibilities. We then used the task {\sc tclean} to make spectral cubes from the residual visibilities, using a range of velocity resolutions (25, 35, 50~\kmps) and values of the robust parameter (-1, -0.5, 0, +0.5, +1). For all cubes, the CO(3--2) line emission was cleaned down to a threshold of 0.5$\sigma$, where $\sigma$ is the per-channel RMS noise. For cleaning, we used a square mask of size $~2.5'' \times 2.5''$, enclosing the CO emission for all frequency planes. Table~\ref{tab:cubes} summarizes the details of the spectral cubes for the different values of the Briggs robust parameter. We chose a robust value of +1 and a velocity resolution of 35~\kmps\ for the final spectral cube used for the kinematic modelling, based on a balance between the per-channel sensitivity and signal-to-noise ratio, and the angular resolution. The final spectral cube has a synthesized beam of $0\farcs32 \times 0\farcs24$, a position angle of $=-74.4^{\circ}$, and an RMS noise of $0.15$~mJy~beam$^{-1}$ at a velocity resolution of 35~\kmps.

\begin{table}[]
    \centering
    \caption{Properties of the spectral cubes, at a channel resolution of 35\,\kmps, with different values of the robust parameter. The columns are (1)~the value of the Briggs robust parameter, (2)~the synthesized beam of the resulting cube, (3)~the RMS noise per channel, in mJy~Beam$^{-1}$, and (4)~the signal-to-noise ratio (S/N) of the peak CO emission in the velocity-integrated CO(3--2) flux density image.}
    \label{tab:cubes}
    \begin{tabular}{cccc}
\hline        
Robust  &  Synthesized beam &   RMS noise &  Peak S/N \\ 

        &                   &    (mJy)         &              \\
\hline 
  +1    &        $0.32'' \times 0.24''$       &       0.15     &        8.1 	\\		
  +0.5  &       $0.27'' \times 0.21''$        &      0.16      &       6.7 	\\	 	
    0   &        $ 0.23'' \times 0.17''$      &        0.19    &         4.7  \\			
  -0.5  &       $ 0.21'' \times 0.17'' $      &       0.23     &        4.7 	\\		
  -1    &       $  0.20'' \times 0.16'' $     &        0.27    &         3.9 	\\		
\hline  
    \end{tabular}
    
\end{table}

\section{HST observations and Data Analysis}

The field of PKS~B1228-113 was observed in 2019 (PID: 15882; PI N.~Kanekar) and 2024 (PID: 17483; PI R.~Dutta) using the Wide Field Camera~3 (WFC3) onboard the Hubble Space Telescope\footnote{All the {\it HST} data used in this paper can be found in MAST: \dataset[10.17909/qmrd-4h86]{http://dx.doi.org/10.17909/qmrd-4h86}.}, with the F105W and F160W filters respectively. For the $z \approx 2.193$ DLA galaxy, this corresponds to rest-frame central wavelengths of $\approx 3000$~\AA\ and $\approx 5000$~\AA, respectively. A single orbit was obtained with each filter, with a larger dither pattern used to achieve cleaner images by dithering over the size of the IR blobs. The F105W observations use the WIDE-7 dither pattern increased by a factor of 3 over the standard size \citep[ISR 2016-14;][]{Anderson:2016}, while the F160W observations used a standard three-point line dither pattern with the size of the dither pattern increased by a factor of 5. 

The individual images were combined with AstroDrizzle \citep{hack20} to a scale of $0\farcs06$~pixel$^{-1}$, and the images astronomically aligned to GAIA DR3 \citep{GAIADR3final} with TweakReg, with an uncertainty of $\approx 0\farcs02$. The effective angular resolution of the HST WFC3 images is $\approx 0\farcs3$, based on Gaussian fits to the point spread function (PSF) of the quasar.

\begin{figure}[t!]
\centering
\includegraphics[width=\textwidth]{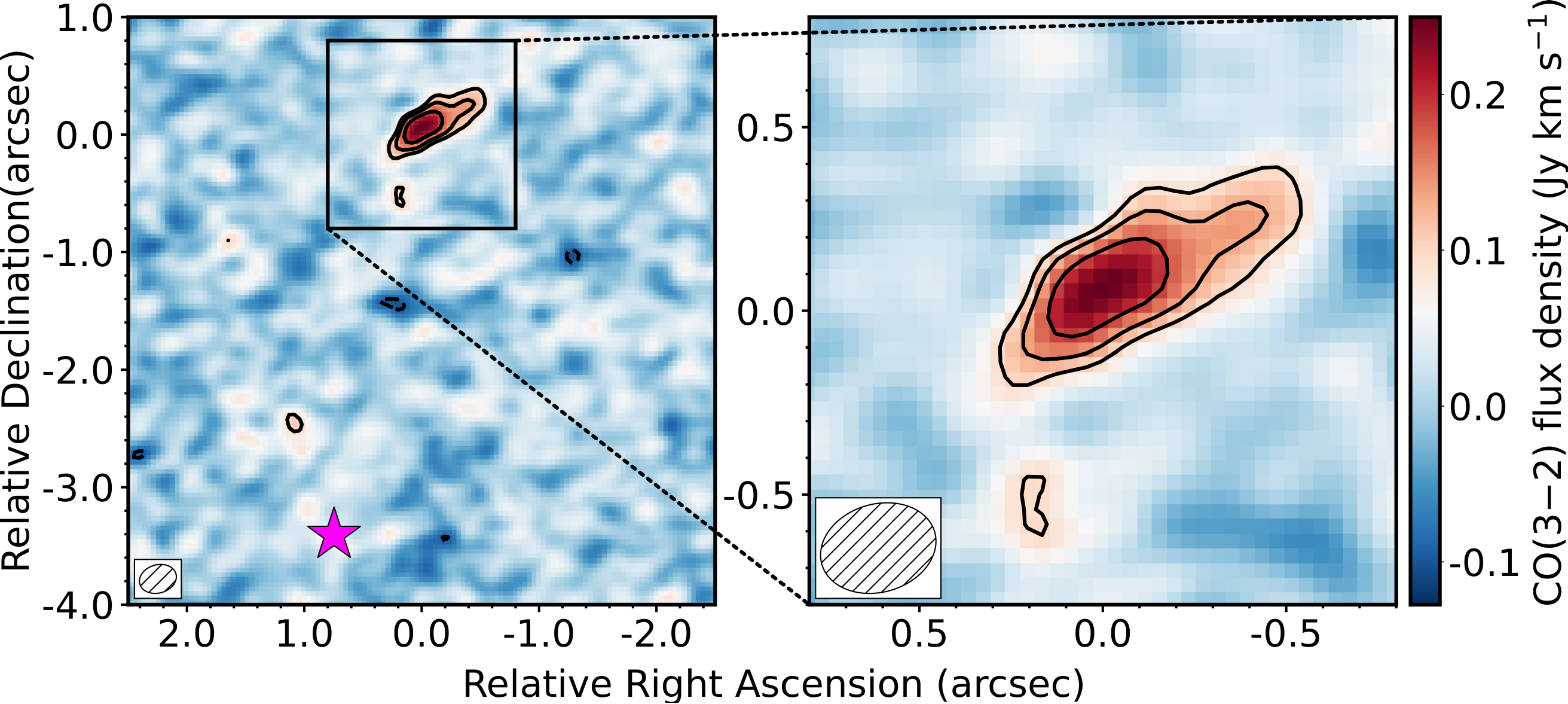}
\caption{The velocity-integrated CO(3--2) image, with contours at $(-3, 3, 4.2, 6) \times \sigma$, where $\sigma$ is the RMS noise on the image. The magenta star indicates the QSO position. The right panel shows a zoom-in on DLA-B1228g. The x-axis and y-axis are right ascension and declination relative to the kinematic centre of the CO(3--2) emission (see Table~\ref{tab:model}), in arcseconds. The synthesized beam is shown in the bottom-left corner of each panel. 
\label{fig:co}}
\end{figure}

\section{Results} 
\label{sec:results}

Figure~\ref{fig:co} shows the velocity-integrated image of the CO(3--2) emission (integrated over the velocity range of $-350$~\kmps\ to $+525$~\kmps) detected in our ALMA spectral cube, at an angular resolution of $\approx 0\farcs32 \times 0\farcs24$ (the corresponding spatial resolution at $z=2.193$ is $2.6 \: {\rm kpc} \times 2.0$~kpc). The QSO position is indicated by the magenta star, $\approx 3\farcs9$ (i.e. $\approx 33$~kpc) to the south of the DLA galaxy \citep{Neeleman18}. The CO(3--2) emission has an integrated line flux density of $(0.73 \pm 0.12)$~Jy~\kmps. This was obtained using the {\sc casa} task {\sc imfit} to fit a single-component 2-D Gaussian model to the image of Figure~\ref{fig:co}; this model was found to yield an acceptable fit, with residuals consistent with noise. The measured integrated CO(3--2) line flux density is consistent with the value of $(0.726 \pm 0.031)$~Jy~\kmps\ obtained from the original low-resolution ALMA CO(3--2) spectrum \citep{Neeleman18,Kanekar20}, indicating that no emission is being resolved out in our high-resolution ALMA cube. The 2-D Gaussian fit to the CO(3--2) emission yields a deconvolved size of $(0\farcs78 \pm 0\farcs15) \times (0\farcs18 \pm 0\farcs07)$, i.e. a physical size of $\approx 6.4 \, {\rm kpc} \times 1.5 \, {\rm kpc}$. No continuum emission is detected at the location of DLA-B1228g in the $0\farcs29 \times 0\farcs22$ resolution continuum image.

\begin{figure}
\centering
\includegraphics[scale=0.6]{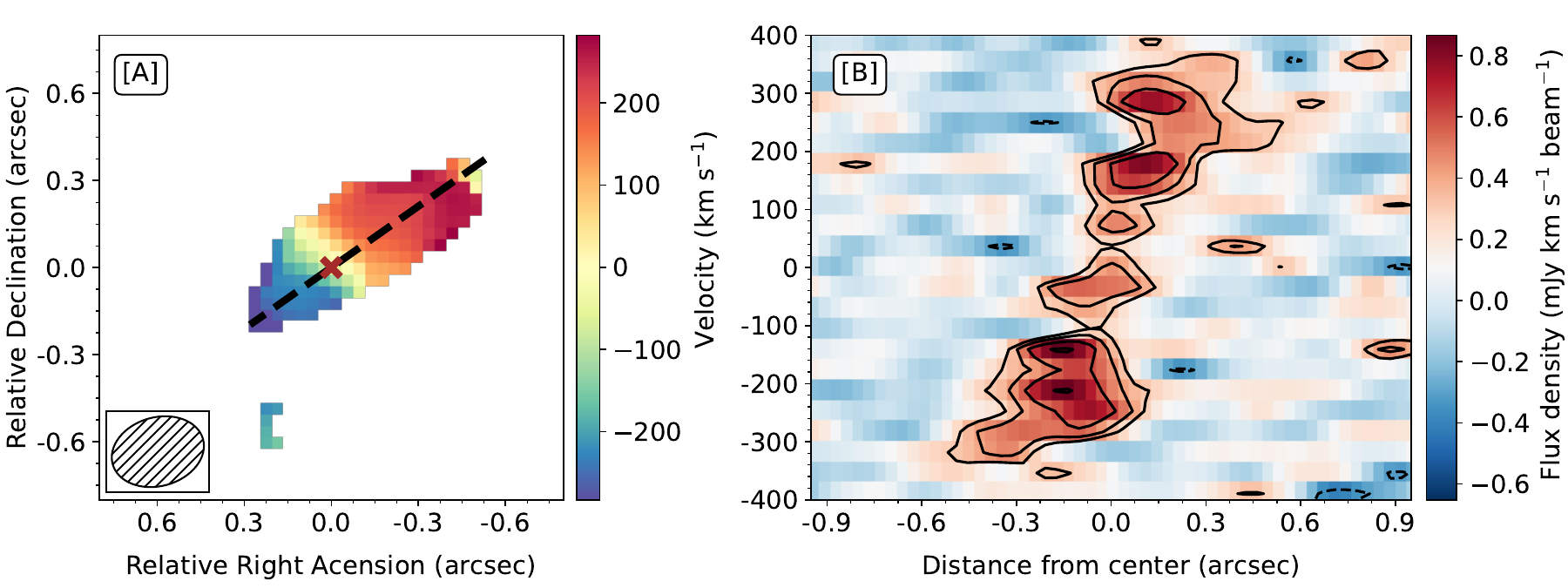}
\caption{[A]~The CO(3--2) mean velocity field, after blanking all spatial locations where the integrated CO(3--2) flux density has $< 3\sigma$ significance. The velocity field is seen to be similar to that of a rotating disk. The red cross indicates the kinematic centre of the CO(3--2) emission (see Table~\ref{tab:model}). [B]~A PV slice along the major axis of the galaxy, indicated by the dashed line in the left panel. The y-axis contains velocity, in \kmps, relative to the DLA redshift, $z=2.1929$, while the x-axis is distance from the the peak of the CO(3--2) emission (see Table~\ref{tab:model}), in arcseconds. The flux density contours start at $\pm2\sigma$ and increase in steps of $\sqrt{2}$, where $\sigma$ is the RMS noise per $35$~\kmps\ channel. 
\label{fig:co-velocity}}
\end{figure}

Figure~\ref{fig:co-velocity}[A] shows the mean CO(3--2) velocity field which was obtained by fitting a Gaussian function to the spectrum at each pixel of the cube, and then using the mean value of the function at each pixel, via the fitting package {\sc qubefit} \citep{Neeleman21}; the image has been blanked at spatial locations where the integrated CO(3--2) flux density has $<3\sigma$ significance. The velocity field appears consistent with that of a rotating disk, as also suggested by the double-horned profile seen in the high-resolution ALMA spectrum. Figure~\ref{fig:co-velocity}[B] shows a position-velocity (PV) slice taken along the major axis\footnote{We note that the major axis was obtained from the kinematic modelling presented in Section~\ref{sec:dis}. Similar results are obtained on taking a PV slice along the major axis of the 2D Gaussian fit to the integrated CO(3--2) intensity.} of the galaxy disk (indicated by the dashed black line in Figure~\ref{fig:co-velocity}[A]). The velocities along the major axis are seen to increase with increasing distance from the galaxy centre, indicative of emission from a rotating disk galaxy. However, we note that the current ALMA CO(3--2) image has intermediate angular resolution and sensitivity, with only $2-3$ synthesized beams across the CO-emitting region. In such situations, it is possible for a merger system to show a velocity field similar to that of a rotating disk \citep[e.g.][]{Rizzo22}. In Section~\ref{sec:disk?}, we will discuss the possibility that DLA-B1228g might be a merger system, rather than a disk galaxy.

\begin{figure}[t!]
\centering
\includegraphics[scale=0.15]{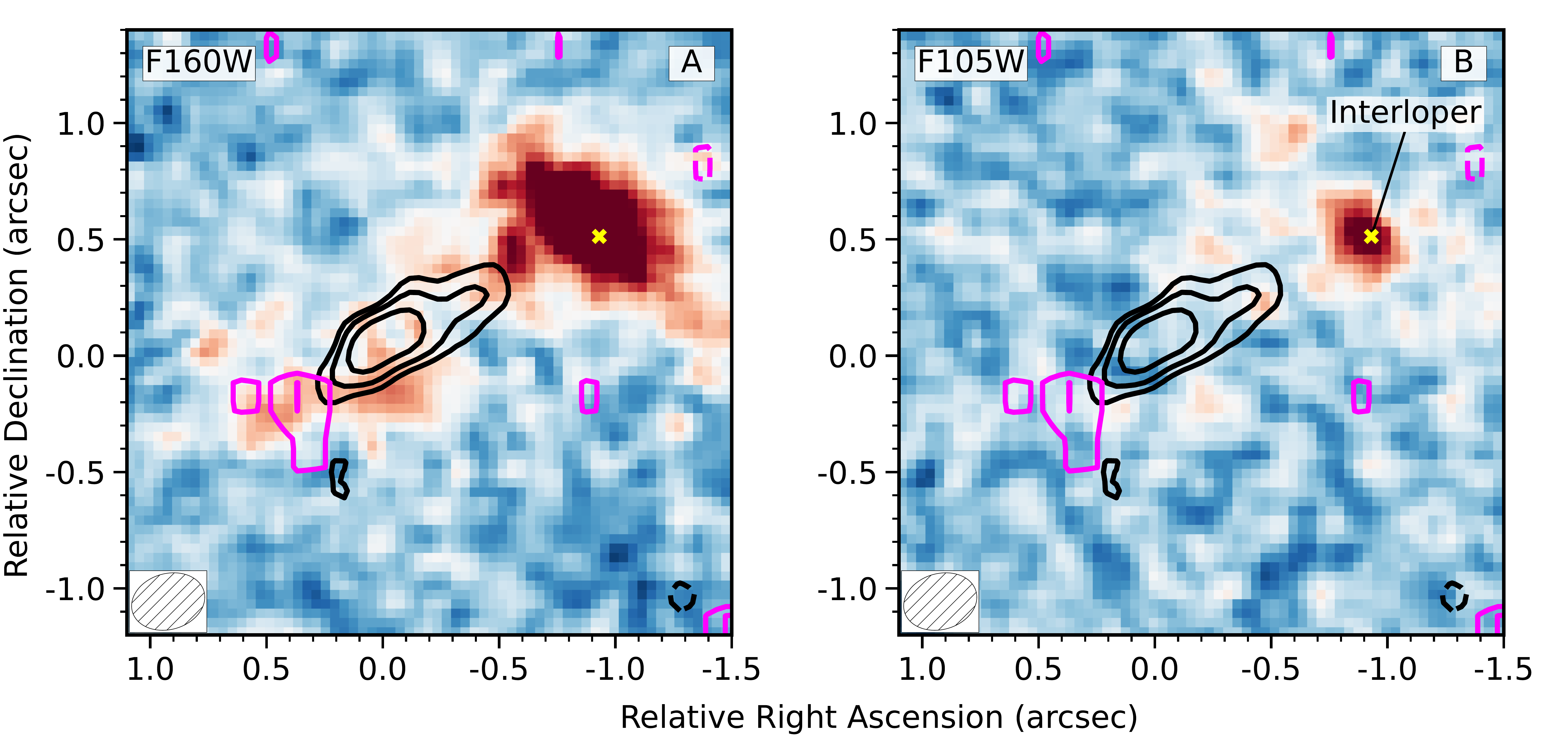}
\caption{ALMA CO(3--2) integrated line flux density image (black contours; both panels) overlaid on [A]~the HST WFC3 F160W image (colourscale) and [B]~the HST WFC3 F105W image (colourscale). The images also show the VLT SINFONI integrated H$\alpha$ intensity image (magenta contours starting at $\pm 3\sigma$ and increasing in steps of $\sqrt{2}$). The x-axis and y-axis are, respectively, right ascension and declination, relative to the kinematic centre of the CO(3--2) emission (see Table~\ref{tab:model}). The inset on the bottom left shows the ALMA synthesized beam. In both panels, the yellow cross indicates the  possible NUV counterpart of DLA-B1228g identified in the HST-WFC3 F105W image by \citet{Kaur22b}. The new ALMA CO(3--2) image shows that this object is clearly offset from the CO emission, and is thus unlikely to be related to DLA-B1228g. See main text for discussion.  
\label{fig:overlay}}
\end{figure}

Fig~\ref{fig:overlay}[A] shows the HST-WFC3 F160W image (in colourscale) overlaid on the ALMA CO(3--2) integrated line flux density image (in black contours). The F160W image shows extended emission (detected at $\approx 8\sigma$ significance) that is approximately coincident with the location of the CO(3--2) emission. The flux-weighted centroid of the HST-WFC3 F160W emission is at RA=12h30m55.510s, Dec.=-11$^\circ$39$\arcmin$06.48$\arcsec$, at an offset of $\approx 0\farcs 2$ (i.e. $\approx 1.5$~kpc) from the kinematic centre of the CO(3--2) emission (see Table~\ref{tab:model} and Section~\ref{sec:dis}). The source has an F160W AB magnitude of $24.5 \pm0.1$, measured over an elliptical aperture with the Kron radius \citep{Kron80}. We find that the HST F160W emission, corresponding to rest-frame 5000\,$\AA$, appears clumpy in nature, with no evidence of a disk-like morphology.

\section{Discussion} \label{sec:dis}

\subsection{Stellar properties of DLA-B1228g}

\citet{Kaur22b} presented an HST-WFC3 image of the field of DLA-B1228g in the F105W filter, corresponding to rest-frame near-ultraviolet (NUV) wavelengths at the galaxy redshift of $z \approx 2.193$. This yielded the apparent detection of the stellar emission of DLA-B1228g, at a position consistent with that of the CO emission in the low-resolution ALMA CO(3--2) and VLA CO(1--0) images \citep{Neeleman18,Kaur22b}. Figure~\ref{fig:overlay}[B] shows the new high-resolution ALMA CO(3--2) image (black contours) overlaid on the HST-WFC3 F105W image (colourscale) and a Very Large Telescope (VLT) SINFONI H$\alpha$ image \citep[magenta contours; ][]{Neeleman18}. The object identified in the HST-WFC3 F105W image by \citet{Kaur22b} is at RA=$12^h30^m55.44^s$,  Dec.=$-11^\circ39\arcmin05.87\arcsec$ (shown by the yellow cross in Figure~\ref{fig:overlay}[B]), and is offset by $\approx 1.02\arcsec$, i.e. $\approx 8.4$~kpc, from the peak of the CO(3--2) emission. 
This object is also clearly detected in our new F160W image of Fig.~\ref{fig:overlay}[A]. The offset between this HST emission feature and the CO(3--2) emission is much larger than the typical spatial offsets of $\lesssim 4$~kpc that have been observed between the stellar and the gas emission in high-$z$ dusty star-forming galaxies \citep[e.g.][]{Hodge15,Calistrorivera18,Bik23}. Further, this HST-WFC3 object shows no H$\alpha$ emission at $z \approx 2.193$ in the VLT-SINFONI image, with an upper limit of $\approx 0.9 \, \Msun$~yr$^{-1}$ on its star-formation rate \citep[SFR;][]{Peroux11}. This, along with the large spatial offset between the optical and the CO emission, indicates that the HST-WFC3 105W source is unlikely to be directly associated with the CO-emitting DLA galaxy.

Of course, as noted in the previous section, we detect extended emission in the new HST-WFC3 F160W image of Fig.~\ref{fig:overlay}[A] which is nearly spatially coincident (offset $\approx1.4$~kpc) with the CO(3--2) emission, and which is not seen in the HST-WFC3 F105W image. This extended emission, at a rest-frame wavelength of $\approx 5000$~\AA\ at $z \approx 2.193$, appears to be the stellar counterpart of the CO(3--2) emission. Dust obscuration, perhaps allied with the Balmer break, is likely to be the main cause of the non-detection of this emission in the F105W image. The F160W emission area is 112~pixels, corresponding to a circular radius of 0.2~arcsec. The $2\sigma$ upper limit on the HST-WFC3 F105W non-detection, at a rest-frame wavelength of $\approx 3000$~\AA\, over this area is AB mag of 26.5. This yields an upper limit on the unobscured SFR of $\approx 1.3~\Msun$~yr$^{-1}$, assuming a Chabrier initial mass function \citep{Chabrier03} and using the local relation between NUV luminosity and the SFR \citep{Kennicutt12}. 
 This is significantly lower than the total SFR of the galaxy measured from the sub-mm continuum, $\approx 100 \, \Msun$~yr$^{-1}$ \citep{Neeleman18,Klitsch22}, indicating a high dust obscuration in the DLA galaxy.

Finally, the two panels of Figure~\ref{fig:overlay} show the VLT SINFONI H$\alpha$ image, which displays emission \citep[with $> 4\sigma$ statistical significance; ][]{Neeleman18} at the south-eastern edge of the CO(3--2) emission. Figure~3 of \citet{Neeleman18} shows that the putative H$\alpha$ emission arises at negative velocities, in good agreement with the CO velocities on the south-eastern side of the disk. It thus appears that the H$\alpha$ emission is detected only from one side of the galaxy.
Interestingly, the new HST-WFC3 F160W image shows rest-frame $\approx 5000$~\AA\ emission at the location of the VLT-SINFONI H$\alpha$ emission. However, H$\alpha$ emission is not detected from the peak of the F160W emission, closest to the centre of DLA-B1228g. This suggests that the brightest F160W emission may arise from an older stellar population, that is not currently undergoing active star formation.

\begin{deluxetable*}{lc}
\tabletypesize{\small}
\tablewidth{5pt} 
\tablecaption{ Results for the kinematic modelling of DLA-B1228g\label{tab:model}}
\tablehead{\hspace{1in}Parameter estimates for the rotating thin-disk model} 
\startdata 
Right Ascension & $12^h30^m55.505(\pm1)^s$\\
Declination &  $-11^\circ39\arcmin06.3870(^{+5}_{-2})\arcsec$ \\
Galaxy redshift & $2.19326 \pm 0.00007$ \\
Inclination (\textit{i}, in deg)    & $66.4\pm2.2$  \\
Position angle (PA, in deg)  & $305.2\pm1.4$  \\
Peak intensity ($I_0$, in mJy/beam) & $3.9^{+0.7}_{-0.5}$ \\
Scale length (R$\rm_d$, in kpc) & $1.8\pm0.1$ \\
Rotation Velocity ($v_{rot}$, in \kmps)       & $328\pm7$ \\  
Velocity Dispersion ($\sigma_{v}$, in \kmps) & $62\pm7$\\
\enddata
\tablecomments{The coordinates indicate the kinematic center of the CO(3--2) emission, obtained from the thin disk modelling of DLA-B1228g.} 
\end{deluxetable*}

\subsection{Kinematic modelling of DLA-B1228g}

\begin{figure}[ht!]
\plotone{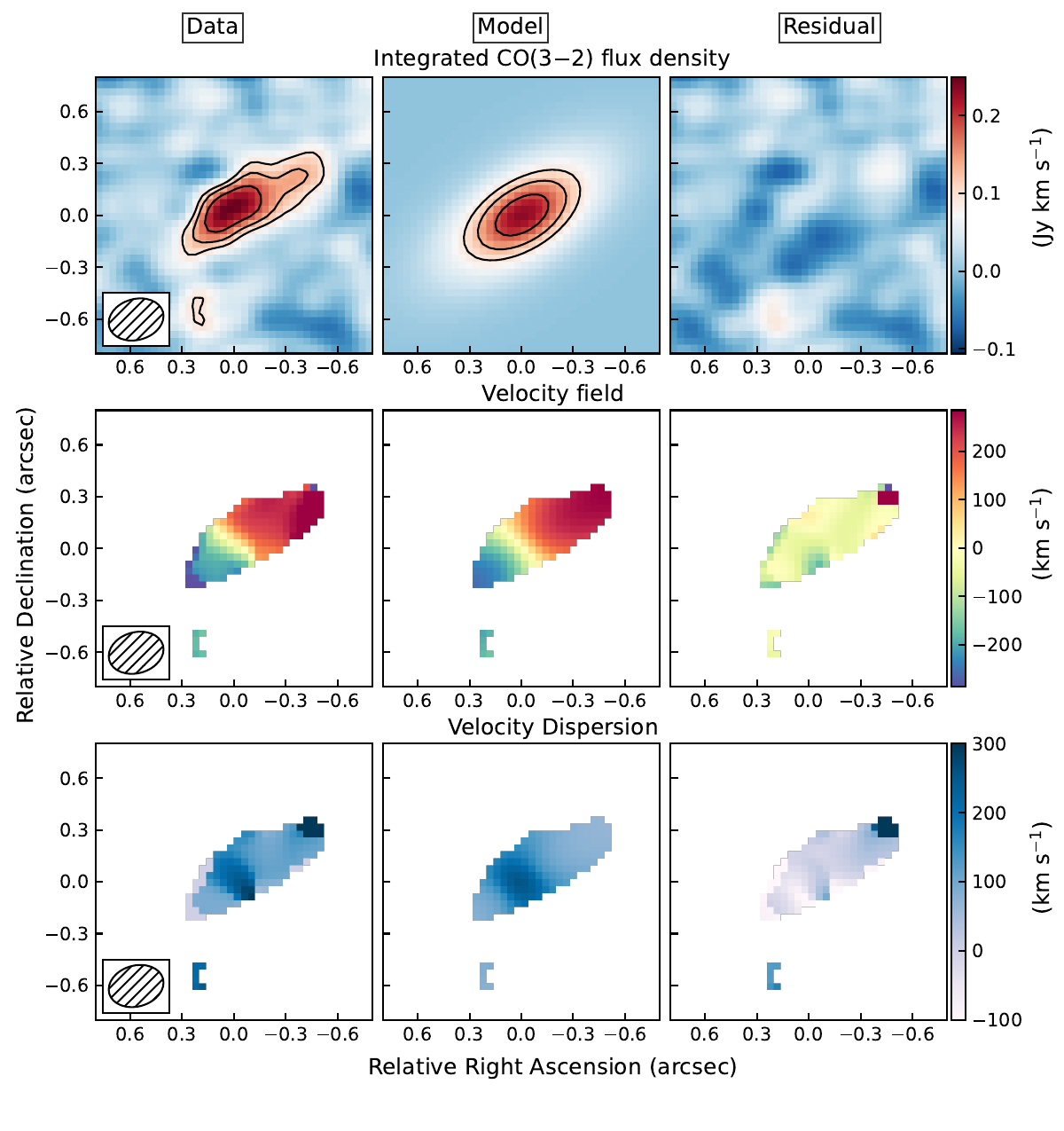}
\caption{The three rows show the CO(3--2) integrated line flux density (Jy~\kmps), the CO(3--2) velocity field (\kmps), and the CO(3--2) velocity dispersion (\kmps), while the three columns show the ALMA image, the best-fit rotating-disk model image, and the residual image obtained after subtracting the model image from the corresponding ALMA image. The x-axis and y-axis are, respectively, right ascension and declination (in arcseconds), relative to the peak of the CO(3--2) emission measured from the best-fit rotating-disk model (see Table~\ref{tab:model}). In the top left panel, the contours start at $\pm3\sigma$, and increase in steps of $\sqrt{2}$; negative features are shown with dashed contours. We note that the top right panel, showing the residual integrated line flux density image, does not contain any residual emission with $\geq 3\sigma$ or $\leq -3\sigma$ significance, and hence does not have any marked contours. The images of the middle and bottom panels have been blanked at spatial locations where the integrated CO(3--2) flux density has $< 3\sigma$ significance. The inset in the left panels shows the ALMA synthesized beam.   
\label{fig:moment}}
\end{figure}

The CO(3--2) velocity field of DLA-B1228g and the PV slice along the galaxy major axis both indicate that the galaxy is likely to be a rotating disk. We used the Bayesian Markov Chain Monte Carlo (MCMC)-based Python fitting package {\sc qubefit} \citep{Neeleman21} to model the kinematics of the CO(3--2) emission. The fitting routine convolves a user-defined model with the synthesized beam of the observations and then compares the resultant model cube to the actual spectral cube, using a defined likelihood function. It then maximizes the likelihood function, following an MCMC approach in order to estimate the model parameters and their uncertainties. We assumed a rotating infinitesimally-thin exponential disk model for DLA-B1228g, in which the intensity $I$ of the CO emission at a radius $r$ is given by $I(r) = I_0 e^{-r/{r_d}}$, where $I_0$ is the central intensity and $r_d$ is the disk scale length. The model is described by the following nine parameters: the coordinates of the centre of the CO emission, the galaxy redshift, the galaxy inclination (\textit{i}), the position angle (PA) of the major axis, the rotation velocity ($v_{rot}$), the velocity dispersion ($\sigma_{v}$), the disk scale length ($r_d$), and the peak intensity of the CO(3--2) emission ($I_0$).  For simplicity, we assume that both the rotational velocity and the velocity dispersion are constant across the disk. Finally, we used 400,000 MCMC chains to estimate the probability distribution functions of the nine parameters of the model.

The results of the kinematic modelling with {\sc qubefit} are shown in Figure~\ref{fig:moment}, where the top row shows the integrated CO(3--2) line flux density, the middle row shows the CO(3--2) velocity field, and the bottom row shows the CO(3--2) velocity dispersion. The three columns contain the data, the model, and the residuals after subtracting the model from the data. It is clear from the top right panel that no residual CO(3--2) emission is detected at a statistically significant ($\geq3\sigma$) level, implying that the assumed thin rotating disk model provides a good fit to the data. The reduced chi-square of the fit is 1.2, consistent with a good fit.

The best-fit parameters of the thin-disk model are listed in Table~\ref{tab:model}; note that the errors have been increased appropriately to account for the fact that the reduced chi-square of the fit is slightly higher than unity. We obtain a rotation velocity of $v_{rot} = 328\pm7$~\kmps, and a velocity dispersion of $\sigma_{v} = 62\pm7$~\kmps. 

The ratio of rotation velocity to velocity dispersion provides information on whether a galaxy's kinematics is dominated by rotation ($v_{rot}/\sigma_v > 1$) or turbulence ($v_{rot}/\sigma_v < 1$). For DLA-B1228g, the kinematic modelling yields the high ratio $v_{rot}/\sigma_v = 5.3\pm0.6$, indicating that the kinematics is dominated by rotation. This is similar to values of $v_{rot}/\sigma_v$ that have recently been obtained for emission-selected galaxies at similar redshifts, when cold gas tracers such as CO, \cii, or C{\sc i} emission are used to map the kinematics \citep[e.g.][]{Lelli23,Rizzo23}

In passing, we note that we used a simple axisymmetric exponential disk model, with a constant rotational velocity and velocity dispersion, for the kinematic modelling as the signal-to-noise ratio of the detected CO(3--2) emission is not very high. Given that this model already provides a good fit, more complicated models (with more parameters) are also likely to yield good fits to the observed CO(3--2) emission.

\subsection{The dynamical mass of DLA-B1228g} 

Our kinematic modelling of the CO(3--2) emission of DLA-B1228g allows us to estimate the dynamical mass of the galaxy, following standard procedures. For a spherically symmetric mass distribution, the dynamical mass enclosed within a radius $R$ is given by 
$M_{dyn} (R) = v^2 \times R/G = 2.33 \times 10^5 \times v^2 \times R \, \Msun$ \citep[e.g.][]{Neeleman21},
where $v$ is the circular velocity, in \kmps, and $R$ is in kpc.

The circular velocity is related to the rotation velocity and the velocity dispersion by $v_{circ}^2 = (v_{rot}^2 + \eta \sigma_v^2)$, where the value of $\eta$ lies in the range $\approx 2-6$ depending on the mass distribution and the galaxy kinematics \citep[e.g.][]{Epinat09,Burkert10}. Since DLA-B1228g is a rotation-dominated system, with $v_{rot}/\sigma_v \approx 5.3$, we will neglect the velocity dispersion in the above expression, and will assume $v_{circ} \approx v_{rot} = 328 \pm 7$~\kmps.

Finally, following \citet{Neeleman21}, we estimate the dynamical mass of DLA-B1228g within a radius equal to twice the effective radius ($R_e$) of the CO(3--2) emission. For an exponential disk, this radius contains $\approx 85$\% of the total emission. Further, for such a disk, the effective radius (i.e. the half-light radius) is equal to $1.678 \times r_d$, where $r_d$ is the disk scale length. For DLA-B1228g, $r_d = 1.8 \pm 0.1$~kpc, implying an effective radius of $R_e = 3.02 \pm 0.17$~kpc. We note that the effective diameter is similar to the estimate of the deconvolved size of the CO(3--2) emission, 6.4~kpc, as measured from the 2-D Gaussian model. 
 The radius that we will use to estimate the dynamical mass is then $R = 2R_e = 6.04 \pm 0.34$~kpc. 

We obtain a dynamical mass of $(1.5 \pm 0.1) \times 10^{11} \, \Msun$ for DLA-B1228g, within the radius $R = 6.04$~kpc. We emphasize that the quoted error only includes the error from the fit. However, the uncertainties in the estimate are dominated by the assumption of a spherical mass distribution. If the actual mass distribution is intermediate between a thin disk and a sphere (e.g. if the baryons are mostly in a disk and the dark matter in a sphere), this estimate would be higher than the true dynamical mass, by $\approx 20$\% \citep[e.g.][]{Neeleman21}.

The dynamical mass estimate includes contributions from the dark matter and the baryons ($\rm M_{bary} \approx \Mmol+M_{HI}+\Ms$) of the galaxy, within a radius of $2R_e$. While the baryonic mass of high-$z$ galaxies has been shown to be dominated by the atomic gas \citep[e.g.][]{Chowdhury22c}, the latter is extended on scales of tens of kpc \citep[e.g.][]{Chowdhury22b}. The baryonic mass of the inner disk is thus likely to be mainly contributed by the molecular gas and the stars. For DLA-B1228g, \citet{Kaur22b} used VLA CO(1--0) spectroscopy to obtain a molecular gas mass of $\Mmol = (1.0 \pm 0.2) \times 10^{11}\times(\alpha_{\rm CO}/4.36)\ M_\odot$. We do not currently have an estimate of the stellar mass of DLA-B1228g. The total baryonic mass within $2R_e$ is thus $\gtrsim 1.0 \times 10^{11}\, \Msun$, using the molecular gas mass estimate alone. This is comparable to the dynamical mass estimate of $\approx 1.5 \times 10^{11} \, \Msun$ (especially given that the dynamical mass is likely to be over-estimated by $\approx 20$\%). We thus find that the central regions of DLA-B1228g are baryon-dominated, as has also been found for emission-selected galaxies at similar redshifts \citep[e.g,][]{Genzel17,Genzel20}

\begin{figure}[t!]
\centering
\includegraphics[scale=0.13]{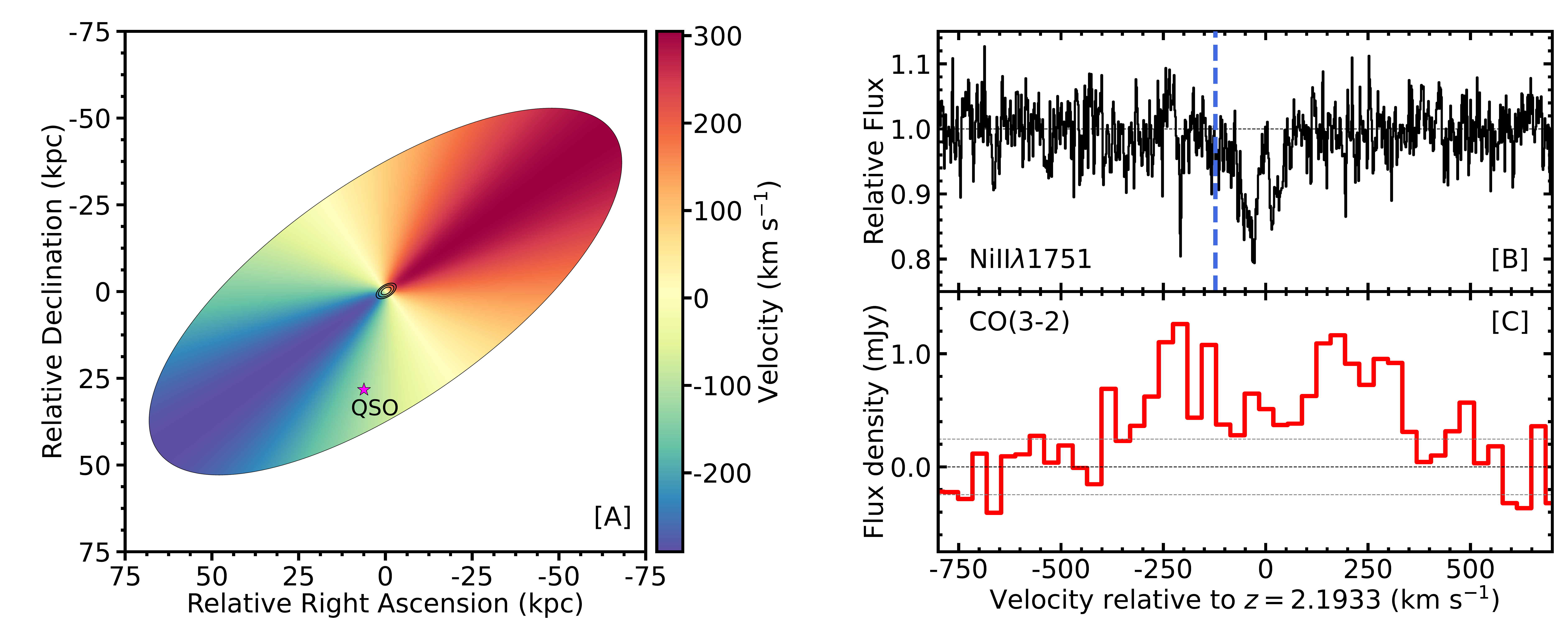}
\caption{The left panel [A] shows the expected \hi\ disk of DLA-B1228g (large ellipse) and its expected velocity field, based on the velocity field of the ALMA CO(3--2) emission (inner ellipse) and an assumed flat rotation curve (see main text for a description of the model for the \hi\ disk). The magenta star indicates the location of the sightline towards the background QSO. 
The x and y axes are in kpc, relative to the kinematic centre of the CO(3--2) emission (see Table~\ref{tab:model}). The top right panel [B]~shows the VLT-UVES Ni{\sc ii}$\lambda$1751\AA\ absorption profile from the $z\approx 2.1929$ DLA, while the bottom right panel~[C] shows the integrated ALMA CO(3--2) flux density spectrum of DLA-B1228g, with the dashed line showing the $\pm1\sigma$ error, where $\sigma$ is the rms noise per $35$ \kmps\ velocity resolution. The vertical blue line in panel~[B] shows the expected absorption velocity if the absorption arises in an extended \hi\ disk with a flat rotation curve (see main text for discussion). In both the right panels, the x-axis is velocity, in \kmps, relative to the redshift of DLA-B1228g ($z = 2.1933$; see Table~\ref{tab:model}).    
\label{fig:diskmodel}}
\end{figure}

\subsection{The $z \approx 2.1929$ DLA: Disk or CGM?}

An important open question in studies of galaxy evolution is whether the high \hi\ column densities characteristic of DLAs arise primarily in the disks of high-$z$ galaxies (i.e. in their ISM) or are also commonly found in their CGM. Indeed, one of the main reasons for the choice of the DLA-defining threshold column density of $2 \times 10^{20}$~cm$^{-2}$ is that such high \hi\ column densities are characteristic of galaxy disks in the local Universe and thus provide a signpost for the presence of a gas-rich galaxy along the QSO sightline \citep{Wolfe86}. 

Very interestingly, the \hi-absorption-selected galaxies associated with DLAs at $z \gtrsim 4$ have impact parameters of $\approx 15-60$~kpc \citep{Neeleman17,Neeleman19}, with no evidence of a closer companion galaxy found in HST imaging \citep{Kaur21}. While the number of identified \hi-absorption-selected galaxies at $z \gtrsim 4$ remains small today, the observational data suggest that DLAs at these redshifts might arise primarily in the CGM.

Conversely, at lower redshifts, $z \approx 2$, most of the presently identified \hi-absorption-selected galaxies have lower impact parameters, $\lesssim 15$~kpc \citep[e.g.][]{Krogager17,Neeleman18,Kanekar20,Kaur22b}, consistent with the DLA arising in the ISM of the associated galaxy. We note that some of the optical techniques \citep[e.g.][]{Moller04,Fynbo10,Wang15,Krogager17} are biased towards identifying DLA galaxies at low impact parameters. 

In this context, the large impact parameter ($\approx 33$~kpc, similar to that of DLA galaxies at $z \approx 4$) of DLA-B1228g to the sightline towards PKS~B1228-113 is unusual. However, this does not immediately imply that the DLA arises in the CGM of the galaxy, because the \hi\ disk of a galaxy is typically far more extended than the stellar or molecular disks. In the local Universe, the \hi\ diameter of a galaxy of a given \hi\ mass can be obtained from the \hi\ mass-size relation \citep[e.g.][]{Broeils97,Wang16}. For example, a galaxy at $z \approx 0$ with an \hi\ mass of $\approx 10^{10} \, \Msun$ has a typical \hi\ diameter (at an \hi\ surface density of $1 \, \Msun$~pc$^{-2}$) of $\approx 60$~kpc \citep{Wang16}. At high redshifts, we do not have measurements of the \hi\ masses of individual galaxies, due to the weakness of the \hii\ emission line. However, recent \hii\ stacking studies have shown that the average \hi\ mass of star-forming galaxies at $z \approx 1.3$ is significantly larger than their molecular gas and stellar masses \citep{Chowdhury22c}. In fact, atomic gas has been shown to make up $\approx 70$\% of the baryonic mass of star-forming galaxies at $z \approx 1.3$ \citep{Chowdhury22c}.
This indicates that the atomic gas mass of DLA-B1228g is likely to be larger than the molecular gas mass, i.e. that the \hi\ mass is $\gtrsim 7.3 \times 10^{10} \times (\alpha_{\rm CO}/4.36) \, \Msun$ (where we have used a factor of 1.36 to convert the atomic gas mass to the \hi\ mass).

Further, \citet{Chowdhury22b} show that the stacked \hii\ emission signal from star-forming galaxies at $z \approx 1$ is spatially resolved for synthesized beams $< 90$~kpc, implying an average \hi\ disk diameter $\gtrsim 50$~kpc for galaxies at $z \approx 1.3$ with an \hi\ mass of $\approx 1.4 \times 10^{10} \, \Msun$. This is consistent with the expected \hi\ mass-size relation at $z \approx 0$ \citep{Wang16}. Assuming that DLA-B1228g follows the local \hi\ mass-size relation and that its \hi\ mass is $\gtrsim 7.3 \times 10^{10} \, \Msun$ yields an \hi\ disk diameter of $\gtrsim 160$~kpc. The solid ellipse in Fig.~\ref{fig:diskmodel}[A] shows the expected \hi\ disk geometry for DLA-B1228g, assuming that the \hi\ disk follows the same geometry as the molecular gas observed in CO emission, for an assumed \hi\ mass of $\approx 7.3 \times 10^{10} \, \Msun$ and the local \hi\ mass-size relation. It is clear in this scenario that the sightline to PKS~B1228-113 is expected to be covered by the \hi\ disk, indicating that it is possible that the damped Lyman-$\alpha$ absorption arises in the \hi\ disk of the galaxy.

Fig.~\ref{fig:diskmodel}[A] also shows the expected \hi\ velocity field of DLA-B1228g, assuming that the \hi\ disk too can be described as a thin exponential disk, with the flat rotation curve of the molecular disk. The expected velocity from the \hi\ disk at the location of the QSO sightline is seen to be $\approx -123$~\kmps. Fig.~\ref{fig:diskmodel}[B] shows the Ni{\sc ii}$\lambda$1751\AA\ absorption line detected in a VLT Ultraviolet Echelle Spectrograph (UVES) spectrum from the $z = 2.1929$ DLA \citep[top panel;][]{Akerman05,Neeleman18} and the integrated ALMA CO(3--2) spectrum (bottom panel; this work). Note that zero velocity in this figure is at the systemic redshift of DLA-B1228g, i.e. at $z = 2.19326$. It is clear that the bulk of the low-ionization metal line absorption lies redward of the expected velocity of $\approx -123$~\kmps\ from the \hi\ disk, extending to a velocity of $\approx +100$~\kmps. A simple thin disk model thus fails to explain the observed low-ionization metal line absorption in the $z \approx 2.1929$ DLA. Scenarios that explain the observed velocity structure of the low-ionization metal lines include absorption in outflowing or accreting gas, in gas clumps in the CGM, or in a warp or flare in the outer disk of the galaxy \citep[e.g.][]{Bouche13}.

\subsection{Can the CO Kinematics be explained by a Major Merger?} 
\label{sec:disk?}

Our kinematic modelling of the CO(3--2) emission has shown that a rotating disk model provides a good fit to the present CO data. However, as emphasized by \citet{Rizzo22}, it is possible for the velocity field of a merger system to appear to be a rotating disk, for observations (such as the present one) with intermediate angular resolution and sensitivity. We hence consider the possibility that DLA-B1228g is actually a merger system, rather than a disk galaxy. 

At the outset, we note that the apparently clumpy nature of the HST F160W emission does appear suggestive of a merger. Further, the kinematic centre obtained from the modelling is offset from the physical centre of the CO emission. Conversely, the double-peaked CO(3--2) emission spectrum of Fig.~\ref{fig:diskmodel}[C] is symmetric, which would be surprising for a merger. Further, the merger would have to be of two galaxies of similar CO(3--2) line flux densities and velocity widths, i.e. of approximately similar molecular gas masses, implying a major merger. One of the putative merging galaxies would then give rise to the blueward CO emission of Fig.~\ref{fig:co-velocity}, and the other, the redward CO emission, with a physical separation of $\approx 3$~kpc and a velocity separation of $\approx 500$~\kmps. Such a major merger would be very likely to trigger massive star formation, producing strong H$\alpha$ emission \citep[e.g,][]{alaghband-zadeh12}. However, (weak) H$\alpha$ emission is seen from only one of the regions showing F160W emission, with no H$\alpha$ emission seen from the second region with the F160W emission, closer to the centre of the CO emission. This too would be very surprising for a major merger. We note that dust obscuration would affect the F160W rest-frame 5000\,\AA\ emission more than the H$\alpha$ emission, and thus cannot explain the non-detection of H$\alpha$ emission from the region with F160W emission. In addition, the velocity dispersions of the blue and red velocity components of the CO emission are seen to be very similar in the bottom-left panel of Fig.~\ref{fig:diskmodel}. In the major-merger scenario, this would be surprising as the velocity dispersions of the two merging galaxies would be expected to be uncorrelated. Further, the DLA sightline lies to the south of DLA-B1228g, i.e. on the side of the blueward CO emission. However, Fig.~\ref{fig:diskmodel}[B] shows that the low-ionization metal line absorption arises roughly midway between the velocities of the two CO peaks. This would be surprising in a merger scenario, as one would expect the absorption to arise from gas that is associated with the galaxy producing the blueward CO emission. Finally, combining the SFR estimate of $\approx 100 \, \Msun$~yr$^{-1}$ \citep{Neeleman18,Klitsch22} with the molecular gas mass of $\approx 10^{11} \, \Msun$ \citep{Kaur22c} yields a molecular gas depletion timescale of  $\approx 1$~Gyr for DLA-B1228g, far larger than typical molecular gas depletion timescales of high-redshift starburst galaxies \citep[$\lesssim 0.1$~Gyr; ][]{Tacconi20}. Assuming a significantly lower value of the CO-to-H$_2$ conversion factor, $\alpha_{\rm CO} \approx 1 \, \Msun$~(K~\kmps~pc$^{2}$)$^{-1}$ \citep[e.g.][]{Bolatto13}, would reduce this discrepancy, but would still yield a molecular gas depletion timescale of $\approx 0.25$~Gyr, higher than that in typical major mergers at $z \approx 2$. Overall, it appears very unlikely that DLA-B1228g is a major merger system, although a definitive test of this scenario would require a high-resolution, high-sensitivity CO imaging study \citep[e.g.][]{Rizzo22}.

In passing, we note that the DLA metallicity of $-0.22$ \citep{Akerman05,Neeleman18} appears quite high, given that the DLA sightline has a high impact parameter,  $\approx 33$~kpc. However, studies of metallicity gradients in large samples of massive emission-selected galaxies at $z \approx 2$ have mostly yielded flat or only slightly negative metallicity gradients, out to radii of $\approx 10$~kpc, and even mildly positive gradients in some galaxies \citep[e.g.][]{Wuyts2016,Forster-Schreiber2018}. In the case of absorption-selected galaxies, \citet{Christensen2014} compared emission and absorption metallicities for two DLAs at $z \approx 2.3-2.6$ and their associated galaxies (at impact parameters of $\approx 6.3$~kpc and $\approx 16$~kpc), obtaining flat metallicity gradients in both cases. Overall, the present data on metallicity gradients in both emission-selected massive galaxies and DLA galaxies at $z \approx 2$ do not suggest any inconsistency with the observed metallicity of $-0.22$ at an impact parameter of $\approx 33$~kpc for the $z \approx 2.193$ DLA towards QSO~B1228-113.

\section{Summary} \label{sec:sum}

We have used ALMA in the extended C43-7 configuration to carry out the first high-resolution CO mapping of a high-redshift, \hi-absorption-selected galaxy, DLA-B1228g at $z \approx 2.193$. We find that the CO(3--2) emission from DLA-B1228g is consistent with arising from a rotating disk, with a spatial extent of $\approx 6.4$~kpc. It appears unlikely that the galaxy is a major merger system, although this scenario cannot be formally ruled out by the present data. Kinematic modelling of the CO(3--2) emission as a rotating thin disk, with an exponential surface density profile, yields a rotation velocity of $v_{rot} = 328\pm7$~\kmps\ and a velocity dispersion of $\sigma_{v} = 62\pm7$~\kmps. The high value of the ratio $v_{rot}/\sigma_v$, $5.3 \pm 0.6$, implies that DLA-B1228g is a cold, rotation-dominated disk galaxy. We obtain a dynamical mass of $\lesssim 1.5 
\times 10^{11} \, \Msun$ from the kinematical modelling, within a radius of $\approx 6$~kpc. This is similar to the molecular gas mass inferred from the VLA CO(1--0) emission, indicating that the inner regions of the galaxy are baryon-dominated. The non-detection of DLA-B1228g in an HST-WFC3 F105W image yields an upper limit of $1.3~\Msun$~yr$^{-1}$ on its unobscured SFR, far lower than the total SFR measured from the dust continuum emission, and indicating that the object is a highly dusty galaxy. We present a new HST-WFC3 F160W image that identifies extended stellar emission (at a rest-frame wavelength of $\approx5000$~\AA\ at $z \approx 2.193$) from DLA-B1228g, close to the CO(3--2) emission, and overlapping with the H$\alpha$ emission detected in an earlier VLT-SINFONI image. The DLA galaxy is likely to have a high \hi\ mass, $\gtrsim 7.3 \times 10^{10} \, \Msun$, with the damped Lyman-$\alpha$ absorption arising from an extended \hi\ disk. However, the velocity offset between the metal-line absorption detected in the VLT-UVES absorption spectrum and the expected disk velocity at the QSO location suggests that a simple thin disk model is not sufficient to explain the observed metal-line absorption.

\begin{acknowledgements}
We thank an anonymous referee for detailed comments and suggestions that have improved this paper. BK and NK acknowledge support from the Department of Atomic Energy, under project 12-R\&D-TFR-5.02-0700; NK also acknowledges support from the Science and Engineering Research Board of the Department of Science and Technology via a J. C. Bose Fellowship (JCB/2023/000030). ALMA is a partnership of ESO (representing its member states), NSF (USA) and NINS (Japan), together with NRC (Canada), NSC and ASIAA (Taiwan), and KASI (Republic of Korea), 
in cooperation with the Republic of Chile. The Joint ALMA Observatory is operated by ESO, AUI/NRAO and NAOJ. The National Radio Astronomy Observatory is a facility of the National Science Foundation operated under cooperative agreement by Associated Universities, Inc. The data reported in this paper are available though the ALMA archive (https://almascience.nrao.edu/alma-data/archive) with project code: ADS/JAO.ALMA \#2019.1.00373.S.
This research is based on observations made with the NASA/ESA Hubble Space Telescope obtained from the Space Telescope Science Institute, which is operated by the Association of Universities for Research in Astronomy, Inc., under NASA contract NAS 5–26555. Support for Program numbers 15882 and 17483 was provided through a grant from the STScI under NASA contract NAS5-26555. This material is based upon work supported by the National Science Foundation under Grant No. 2107989.
\end{acknowledgements}

\facilities{ALMA, HST}

\software{Qubefit \citep{Neeleman21}, Astropy \citep{astropy:2018}, CASA \citep{casa22,Hunter23}.}

\bibliography{bibliography}{}
\bibliographystyle{aasjournal}

\end{document}